\documentclass[pra,aps,twocolumn,showpacs]{revtex4}

\usepackage{amsmath}
\usepackage{bm}
\usepackage{graphicx}

\begin{document}

\title{Spontaneous magnetization and structure formation in a spin-1
ferromagnetic Bose-Einstein condensate}

\author{Hiroki Saito}
\author{Masahito Ueda}
\affiliation{Department of Physics, Tokyo Institute of Technology,
Tokyo 152-8551, Japan \\
and CREST, Japan Science and Technology Corporation (JST), Saitama
332-0012, Japan
}

\date{\today}

\begin{abstract}
Motivated by recent experiments involving the non-destructive imaging of
magnetization of a spin-1 ${}^{87}{\rm Rb}$ Bose gas (Higbie {\it et al.},
cond-mat/0502517), we address the question of how the spontaneous
magnetization of a ferromagnetic BEC occurs in a spin-conserving system.
Due to competition between the ferromagnetic interaction and the total
spin conservation, various spin structures such as staggered magnetic
domains, and helical and concentric ring structures are formed, depending
on the geometry of the trapping potential.
\end{abstract}


\maketitle

\section{Introduction}

Bose-Einstein condensates (BECs) of atomic gases with spin degrees of
freedom have revealed a rich variety of static and dynamic properties.
The first spinor BEC was realized in an optical trap by the MIT
group~\cite{Kurn} in a spin-1 ${}^{23}{\rm Na}$ BEC which was found to
have the antiferromagnetic ground state~\cite{Stenger}.
The same group subsequently demonstrated the formation of spin
domains~\cite{Miesner} and quantum tunneling across the spin
domains~\cite{Kurn2}.
Spin exchange dynamics were recently observed in both
spin-1~\cite{Chang} and spin-2~\cite{Schmal,Kuwamoto} ${}^{87}{\rm Rb}$
condensates.
The thermodynamic properties of spinor Bose gas were investigated in
Ref.~\cite{Schmal2}.

In the above experiments, the spinor BECs were observed by destructive
absorption imaging after the Stern-Gerlach spin decomposition parallel to
the magnetic field, and therefore the observed quantity was the population
in each magnetic sublevel.
Recently, Higbie {\it et al.}~\cite{Higbie} performed nondestructive
imaging sensitive to the direction of the magnetization, where the probe
axis was taken to be perpendicular to the magnetic field.
The observed quantity was therefore the magnetization perpendicular to the
magnetic field, and the Larmor precession was observed.
Higbie {\it et al.}~\cite{Higbie} suggested the possibility of the
spontaneous magnetization of a spin-1 ${}^{87}{\rm Rb}$ BEC due to
ferromagnetic interactions.
Motivated by that study, in the present paper, the spinor dynamics in a
ferromagnetic BEC are investigated in order to elucidate how spontaneous
magnetization of the ferromagnetic spin-1 BEC occurs in an isolated system
in which the total spin angular momentum is conserved.

We consider a spin-1 ferromagnetic BEC at absolute zero and assume that
all atoms in the BEC are initially prepared in the $m = 0$ magnetic
sublevel.
Here $m$ denotes the projection of the spin on the quantization axis,
which we choose to be the $z$ axis.
Because of the ferromagnetic interaction, we expect that the magnetization
grows spontaneously.
However, global magnetization is prohibited because the total
magnetization is conserved and should remain zero.
We find that the system develops local magnetic domains of various types,
which depend on the geometry of the trapping potential.
We will show that an elongated BEC in the $m = 0$ magnetic sublevel is
dynamically unstable against the formation of magnetic domains, in which
the directions of the spin in adjacent domains are opposed.
This staggered magnetic domain structure is also dynamically unstable, and
subsequently develops into a helical structure due to the ferromagnetic
interaction.
In a tight pancake-shaped BEC, on the other hand, a concentric ring
structure of the spin is created.

This paper is organized as follows.
Section~\ref{s:formulate} briefly reviews the formulation of the spin-1
BEC.
Section~\ref{s:trap} presents the results of our numerical simulations of
the time evolution of trapped spinor systems.
Section~\ref{s:discuss} discusses the physical origins of the structure
formation of spinor BECs found in Sec.~\ref{s:trap}, and
Sec.~\ref{s:conclusions} provides concluding remarks.

\section{Formulation of the problem}
\label{s:formulate}

The $s$-wave scattering between two spin-1 bosons is characterized by the
total spin of the two colliding bosons, $0$ or $2$, and we denote the
corresponding scattering lengths by $a_0$ and $a_2$.
The mean-field interaction energy can be written as $\int d{\bf r} [ c_0
n^2 + c_1 \bm{F}^2 ] / 2$, where the coefficients are given by~\cite{Ho}
\begin{equation}
c_0 = \frac{4 \pi \hbar^2}{M} \frac{a_0 + 2 a_2}{3}
\end{equation}
and
\begin{equation}
c_1 = \frac{4 \pi \hbar^2}{M} \frac{a_2 - a_0}{3}
\end{equation}
with $M$ being the mass of the boson.
The system is ferromagnetic if $c_1 < 0$, and antiferromagnetic if $c_1 >
0$.
The density of particles is defined by
\begin{equation}
n = \sum_{m = -1}^1 |\psi_{m}|^2,
\end{equation}
and the components of the spin vector are given by $F_x = (F_+ + F_-) /
2$, $F_y = (F_+ - F_-) / (2i)$, and
\begin{equation}
F_z = |\psi_1|^2 - |\psi_{-1}|^2
\end{equation}
with
\begin{equation}
F_+ = F_-^* = \sqrt{2} (\psi_1^* \psi_0 + \psi_0^* \psi_{-1}),
\end{equation}
where $\psi_m$ is the mean-field wave function for the magnetic sublevels
$m = 1, 0$, and $-1$.
Since $c_1$ is negative for the spin-1 ${}^{87}{\rm Rb}$ BEC, its ground
state at a zero magnetic field is ferromagnetic~\cite{Ho,Klausen}.

In the presence of a magnetic field, the energy of an alkali atom is
shifted, primarily due to the electron magnetic moment.
We take this effect into account up to the second order in the magnetic
field.
The linear and quadratic Zeeman energies of the hyperfine spin $f = 1$
state are given by $p = -\mu_{\rm B} B / 2$ and $q = \mu_{\rm B}^2 B^2 /
(4 \Delta_{\rm hf})$, respectively, where $\mu_{\rm B}$ is the Bohr
magneton and $\Delta_{\rm hf}$ is the hyperfine splitting~\cite{Pethick}.
In the case of the $f = 1$ ${}^{87}{\rm Rb}$ BEC, we have $p / (k_{\rm B}
B) \simeq -34$ $\mu{\rm K} / {\rm G}$ and $q / (k_{\rm B} B^2) \simeq 3.5$
${\rm nK} / {\rm G}^2$.
The multicomponent Gross-Pitaevskii (GP) equations are thus given by
\begin{subequations}
\label{GP}
\begin{eqnarray} 
i \hbar \frac{\partial \psi_0}{\partial t} & = & \left(-\frac{\hbar^2}{2 M}
\nabla^2 + V \right) \psi_0 + c_0 n \psi_0 \nonumber \\
& & + \frac{c_1}{\sqrt{2}} \left( F_+ \psi_1 + F_- \psi_{-1} \right),
\label{GP1} \\
i \hbar \frac{\partial \psi_{\pm 1}}{\partial t} & = &
\left(-\frac{\hbar^2}{2 M} \nabla^2 + V \pm p + q \right) \psi_{\pm 1} +
c_0 n \psi_{\pm 1} \nonumber \\
& & + c_1 \left( \frac{1}{\sqrt{2}} F_\mp \psi_0 \pm F_z
\psi_{\pm 1} \right). \label{GP2}
\end{eqnarray}
\end{subequations}

In the following section, we will numerically solve the GP equations
(\ref{GP1}) and (\ref{GP2}) in order to examine how the spontaneous
magnetization occurs under the restriction of the conservation of the
total magnetization.
We will take the scattering lengths of ${}^{87}{\rm Rb}$, $a_0 = 101.8
a_{\rm B}$ and $a_2 = 100.4 a_{\rm B}$, from Ref.~\cite{Kempen}, where
$a_{\rm B}$ is the Bohr radius. 
Following the experiments by Higbie {\it et al.}~\cite{Higbie}, we first
prepare the ground state $\psi_{\rm g}$ to be in the $m = -1$ state by
using the imaginary-time propagation method~\cite{Dalfovo}.
The spin-polarized condensate is then transferred into the $m = 0$ state
as an initial state.
In order to simulate experimental imperfections in this population
transfer, we assume very small populations in the $m = -1$ component
($\psi_{-1}(t = 0) = 0.01 \psi_{\rm g}$) in the initial state.
Thus, the initial state is a state of broken symmetry with very small
magnetizations in the $x$ and $z$ directions, which trigger the
spontaneous magnetization.
If $\psi_{\pm 1}(t = 0)$ are exactly zero, the system develops no
spontaneous magnetization within the framework of the mean field
approximation.
The time evolution of the system is obtained by numerically solving the GP
equations (\ref{GP}) by the finite difference method with the
Crank-Nicholson scheme.

We apply the magnetic field of 54 mG along the trap axis, i.e., in the $z$
direction~\cite{Higbie}, which induces the Larmor precession of the spin
in the $x$-$y$ plane at a frequency of $p / (2\pi \hbar) \simeq 38$ kHz.
Since this frequency is much larger than the characteristic frequencies of
the system's dynamics, we set $p = 0$ in the calculation to avoid the
rapid oscillations.
We note that this procedure does not alter any of the dynamics of the
system other than eliminating the Larmor precession of the spin around the
$z$ axis.

\section{Spin dynamics in trapped systems}
\label{s:trap}

\subsection{Spin dynamics in a cigar-shaped trap}

We will first numerically simulate the experimental situation of
Ref.~\cite{Higbie}.
Although the trap frequencies in the experiment are $(\omega_x, \omega_y,
\omega_z) = 2\pi (150, 400, 4)$ Hz, we assume an axisymmetric trap with
the radial frequency $\omega_\perp$ taken to be equal to the geometric
mean $\omega_\perp = \sqrt{\omega_x \omega_y}$ in order to save the
computational task.
The number of atoms is assumed to be $N = 4 \times 10^6$.

\begin{figure}[tb]
\includegraphics[width=8.4cm]{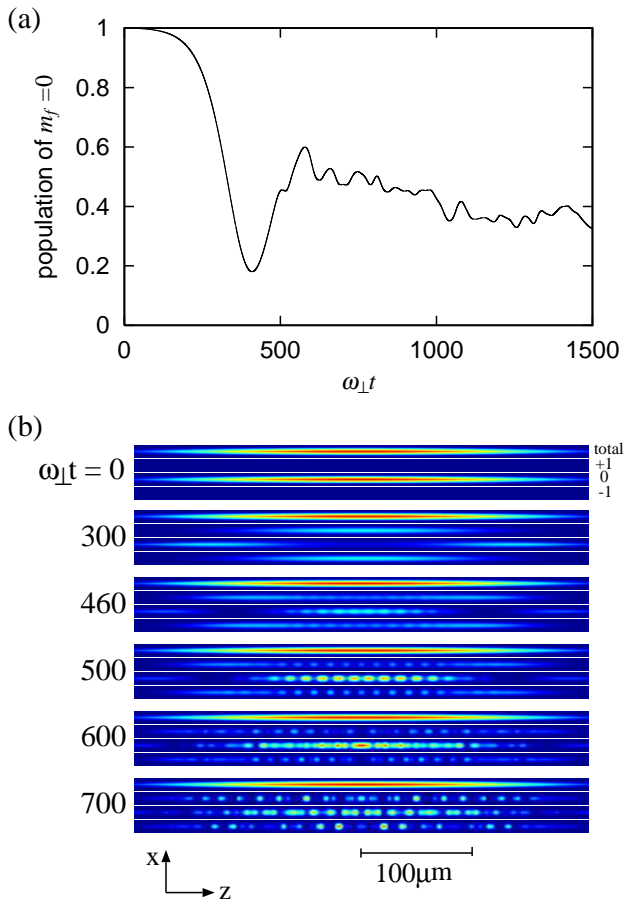}
\caption{
(a) Time evolution of the relative population of the $m = 0$ component,
$\int d{\bf r} |\psi_0|^2 / N$, and (b) the column density distribution
of the total density, $\int dy \sum_{m = -1}^1 |\psi_m|^2$, and those of
the $m = 1$, $0$, and $-1$ components, $\int dy |\psi_m|^2$, from top to
bottom.
The trap frequencies are $\omega_\perp = 2\pi \times 245$ Hz and $\omega_z
= 2\pi \times 4$ Hz.
The magnetic field of 54 mG is applied in the $z$ direction.
The initial state is given by $\psi_1 = 0$, $\psi_0 = \sqrt{1 - 10^{-4}}
\psi_{\rm g}$, and $\psi_{-1} = 0.01 \psi_{\rm g}$, where $\psi_{\rm g}$
is the $m = -1$ ground state.
The size of each strip is $600 \times 16$ in units
of $(\hbar / m \omega_\perp)^{1/2} \simeq 0.69 \mu{\rm m}$.
}
\label{f:evolution}
\end{figure}

Figure~\ref{f:evolution} shows the time evolution of the $m = 0$
population (Fig.~\ref{f:evolution}(a)) and that of the column density of
each spin component (Fig.~\ref{f:evolution}(b)).
The $m = 0$ population first decreases due to the spin-exchange
interaction $0 + 0 \rightarrow 1 + (-1)$.
It should be noted that the central part of the condensate predominantly
converts to the $m = \pm 1$ state (Fig.~\ref{f:evolution} (b),
$\omega_\perp t = 300$) because the spin-exchange interaction proceeds
faster for higher density.
This phenomenon was observed in Ref.~\cite{Kuwamoto}.
The $m = \pm 1$ components then swing back to the $m = 0$ state, and
then the dynamical (modulational) instability sets in at $\omega_\perp t
\simeq 460$.
The time evolution of the $m = 0$ population in Fig.~\ref{f:evolution}
(a) marks the onset of the dynamical instability when the large-amplitude
oscillation changes to a small-amplitude oscillation at $\omega_\perp t
\sim 500$.
After $\omega_\perp t \sim 700$, irregular spin domains appear, as shown
in the bottom panel in Fig.~\ref{f:evolution} (b), and each spin component
as a whole is almost equally populated.
Here, the domain structure is not static but changes in time in a
complex manner.
We note that the total density $n$ (the top image in each set in
Fig.~\ref{f:evolution} (b)) remains almost unchanged, despite the dynamic
evolution of each spin component.
This is due to the fact that the spin-independent interaction energy, $c_0
n / k_{\rm B} \sim 100$ nK, is much larger than the spin-dependent
interaction energy, $c_1 n / k_{\rm B} \sim 1$ nK.

\begin{figure}
\includegraphics[width=8.4cm]{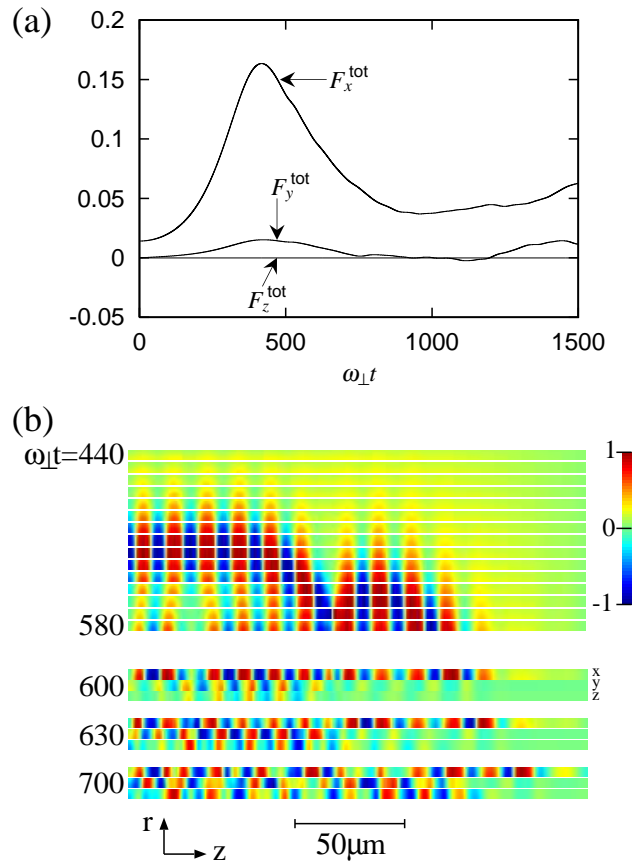}
\caption{
(a) Time evolution of the three components of the total spin per atom,
$\bm{F}^{\rm tot} / N \equiv \int d{\bf r} \bm{F} / N$, and (b) the
radial and axial distributions of the three components of the mean spin
vector, $\bm{F} / n$.
The conditions are the same as those given in
Fig.~\protect\ref{f:evolution}, and the data are taken on the rotating
frame of reference, in which the 38 kHz Larmor precession in the $x$-$y$
plane is eliminated for the sake of clarity.
Each strip in (b) indicates the radial (ordinate) and axial (abscissa)
distribution, where the bottom left corner is the origin ($r = z = 0$).
From $t = 440$ to $580$, only the $F_x / n$ distributions are shown
at every $\omega_\perp \Delta t = 10$, because the other components are
negligibly small.
After $\omega_\perp t = 600$, $F_x / n$, $F_y / n$, and $F_z / n$
are shown from top to bottom.
The size of each strip is $300 \times 6$ in units of $(\hbar / m
\omega_\perp)^{1/2}$.
}
\label{f:spin}
\end{figure}

The time evolutions and spatial dependences of the three components of the
spin vector are shown in Fig.~\ref{f:spin}, where the Larmor precession at
38 kHz is eliminated to avoid the rapid oscillations in the figure.
The time evolution of each components of the total spin per atom
$\bm{F}^{\rm tot} / N \equiv \int d{\bf r} \bm{F} / N$ is shown in
Fig.~\ref{f:spin} (a).
The initial state is taken to have a small $m = -1$ component ($\psi_{-1}
= 0.01 \psi_{\rm g}$, $\psi_{\rm g}$ being the $m = -1$ ground-state wave
function), so that the initial values of the spin are given by $F_x^{\rm
tot} / N \simeq 0.014$, $F_y^{\rm tot} / N = 0$, and $F_z^{\rm tot} / N
\simeq -10^{-4}$.
The $x$ component first increases to $\sim 0.16$ and then decreases.
Due to the quadratic Zeeman effect, the $x$ and $y$ components of the
total spin are not conserved, but the $z$ component is conserved due to
the axisymmetry of the Hamiltonian.

Figure~\ref{f:spin} (b) shows each component of the normalized spin
density $\bm{F} / n$, which corresponds to the mean spin vector per
atom.
Only the $x$ component is shown for $\omega_\perp t \leq 580$ since the
$y$ and $z$ components are negligibly small.
Here, we can clearly see that the staggered spin domain structure is
formed along the $z$ direction at $\omega_\perp t \simeq 500$, due to the
dynamical instability.
The formation of spin domains with staggered magnetization originates from
the conservation of the total magnetization.
Since magnetization of the entire condensate in the same direction is
prohibited due to spin conservation, spontaneous magnetization can occur
only locally at the cost of the kinetic energy of the domain walls.
As mentioned above, the projected total spin on the $x$-$y$ plane is not,
strictly speaking, conserved.
However, since the magnetic field is very weak ($B = 54$ mG), the
nonconserving spin components remain insignificant (see Fig.~\ref{f:spin}
(a)).

The domains are first created around the center of the BEC with $|z|
\lesssim 60 \mu{\rm m}$, since the dynamical instability proceeds faster
in a higher-density region; afterward, the domains gradually extend over
the entire condensate.
In each domain, the length of the mean spin vector is of order of unity,
and the ferromagnetic state is thus formed locally.
At $\omega_\perp t \simeq 600$, the $y$ and $z$ components start to grow,
and the spin domains evolve in a complex manner.
The fact that the growth occurs first in the $x$ component and then in the
other components is due to the initial condition, in which small
magnetization is assumed in the $x$ direction.

\begin{figure}
\includegraphics[width=8.4cm]{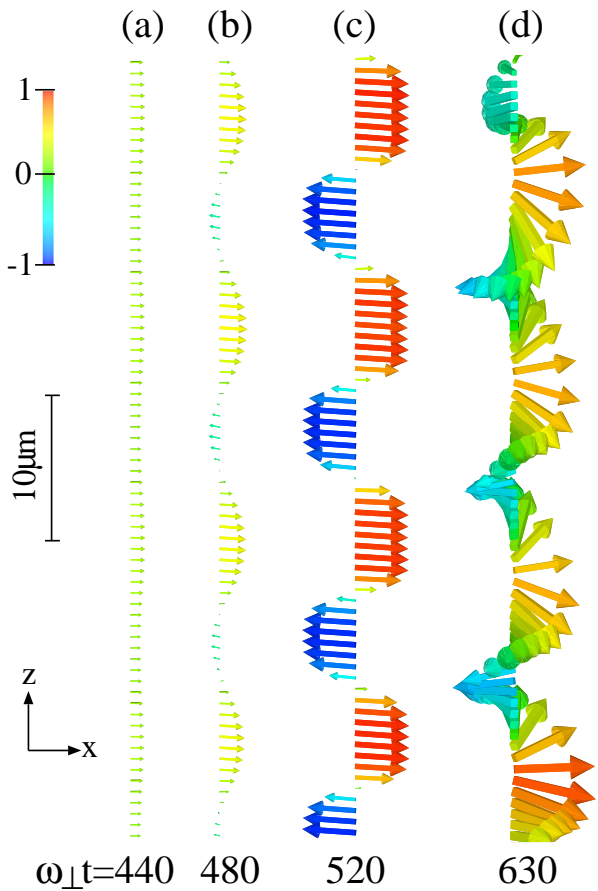}
\caption{
The mean spin vectors $\bm{F} / n$ at $r = 0$ seen from the $-y$
direction, where the vertical axis is the $z$ axis.
The conditions are the same as those given in
Fig.~\protect\ref{f:evolution}.
The length of the spin vector is proportional to $|\bm{F}| / n$, and
the color represents $F_x / n$ according to the gauge shown at the top
left corner.
The spin vector is displayed from $z = 0$ to $z = 54 \mu{\rm m}$ in
(a)-(c), and from $z = 35 \mu{\rm m}$ to $z = 89 \mu{\rm m}$ in (d).
The Larmor precession in the $x$-$y$ plane is eliminated for clarity of
presentation.
}
\label{f:vector}
\end{figure}

The $z$ dependence of the mean spin vectors $\bm{F} / n$ at $r = 0$
is shown in Fig.~\ref{f:vector}.
The staggered magnetic domains grow from $\omega_\perp t = 440$ to $520$
(see Fig.~\ref{f:vector} (a) to (c)).
The spin vectors in the domains in (c) slightly incline in the $z$
direction due to the small initial value of $F_z$.
Figure~\ref{f:vector} (d) shows a snapshot at an early stage of the growth
of the $y$ component.
The growing $y$ component also has a staggered structure, which is shifted
from that of the $x$ component by one quarter of the wavelength.
As a consequence, the spin vectors rotate clockwise or counterclockwise
around the $z$ axis, thereby forming a helical structure, as shown in
Fig.~\ref{f:vector} (d).
Fragmented helical structures having typically one or two cycles of a
helix are also formed in the course of the dynamics after $\omega_\perp t
\sim 700$, but the lifetime of these structures is very short, i.e.,
$\lesssim 10 \mu{\rm s}$.
In contrast, the helical structure observed in Ref.~\cite{Higbie} is not
spontaneously formed, but ``forced'' by an external magnetic field.

The transverse spin distribution, e.g., $F_x$, can be observed by the
nondestructive imaging method in Ref.~\cite{Higbie}.
The staggered domain structure in Fig.~\ref{f:vector} (c) should then look
like commensurate soliton trains, since the optical thickness is high in
the red area with a large $F_x$.
The red and blue areas alternately blink because of the Larmor
precession.
The image of the helical structure moves along the $z$ axis due to the
Larmor precession as observed in Ref.~\cite{Higbie}.

\subsection{Spin dynamics in a pancake-shaped trap}

Next, we will examine the case of a pancake-shaped trap, in which the
axial frequency $\omega_z$ is much larger than both the radial frequency
$\omega_\perp$ and the characteristic frequency of the spin dynamics.
In this case, the spin dynamics occur primarily in the $x$-$y$ plane, in
contrast to the case of the cigar-shaped trap.
We therefore assume that the system can be reduced to two dimensions (2D)
in the $x$-$y$ plane.
The effective strength of the interaction in the quasi-2D system is
obtained by multiplying $c_0$ and $c_1$ by $(8 \pi m \omega_z /
\hbar)^{1/2}$~\cite{Castin}.

\begin{figure}
\includegraphics[width=8.4cm]{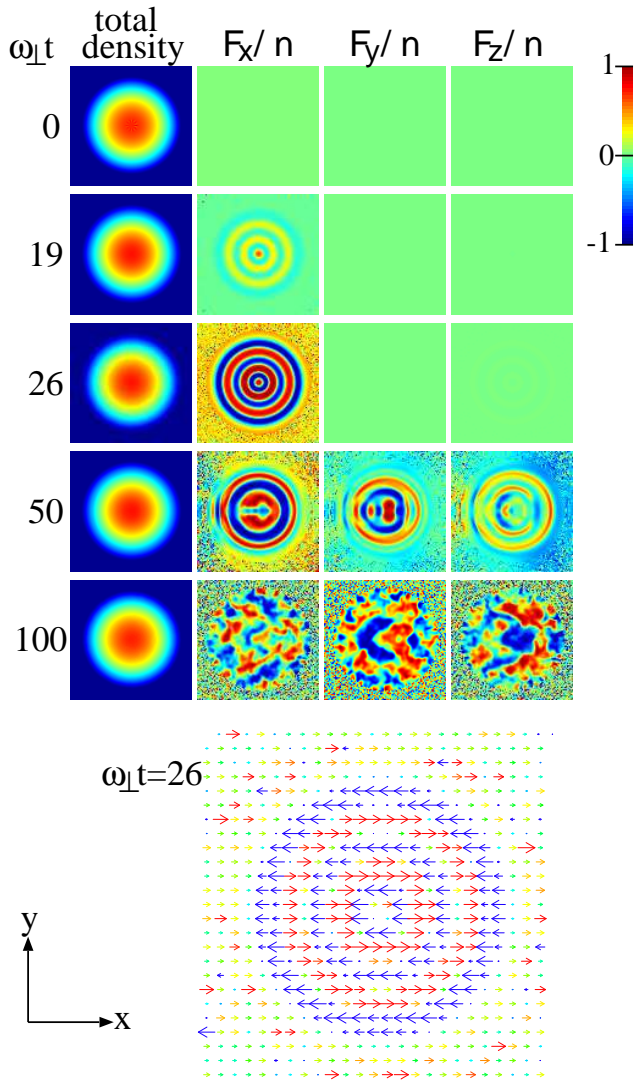}
\caption{
The total density and the three components of the mean spin vector $\bm{F}
/ n$ of the two-dimensional system (from left to right panel), where 
the abscissa and ordinate refer to the $x$ and $y$ coordinates in real
space.
The size of each image is $48 \times 48$ in units of $(\hbar / m
\omega_\perp)^{1/2}$.
The color for the mean spin vector refers to the gauge.
The bottom panel illustrates the direction of the spin at $\omega_\perp t
= 26$, where the color represents $F_x / n$ as shown in
Fig.~\protect\ref{f:vector}.
The Larmor precession in the $x$-$y$ plane is eliminated.
}
\label{f:2d}
\end{figure}

Figure~\ref{f:2d} shows the snapshots of the total density and the $x$,
$y$, $z$ components of the mean spin vector $\bm{F} / n$, where
$\omega_\perp = \omega_z / 20 = 2 \pi \times 10$ Hz, the number of atoms
is $2 \times 10^6$, and the strength of the magnetic field is 54 mG.
The initial state is taken to be $\psi_0 = \sqrt{1 - (0.01)^2} \psi_{\rm
g}$, $\psi_{-1} = 0.01 \psi_{\rm g}$, and $\psi_1 = 0$, where $\psi_{\rm
g}$ is the ground state of the $m = -1$ state obtained by the
imaginary-time method.
We then multiply the initial wave function by a small anisotropic
perturbation $1 + 0.001 \sum_{j = 1}^{10} r^j e^{-r^2 + i j \phi}$ in
order to simulate experimental imperfection in preparing the ground state,
where $r = (x^2 + y^2)^{1/2}$ and $\phi$ is the azimuthal angle.
Despite this anisotropic perturbation, the concentric ring structure of
the spin (Fig.~\ref{f:2d}, $\omega_\perp t = 26$) arises from the dynamical
instability, the origin of which is the competition between the
ferromagnetic interaction and the spin conservation.
The spatial distribution of the directions of the spin vector at
$\omega_\perp t = 26$ is illustrated at the bottom of Fig.~\ref{f:2d}.
The spin vector rotates around the $z$ axis at the Larmor precession
frequency, and hence the spin texture displayed as the red and blue rings
should be observed alternately in time by the imaging method in
Ref.~\cite{Higbie}.
The axisymmetry is spontaneously broken at $\omega_\perp t \simeq 50$ and
the $y$ and $z$ components of the spin vector also grow.
After these events occur, the axisymmetry is completely lost and
complicated spin dynamics emerge as shown in the fifth row ($\omega_\perp
t = 100$) of Fig.~\ref{f:2d}.
Throughout this time evolution, the total density remains almost
unchanged, as in the case of the cigar-shaped geometry.

\section{Structure formation of spinor condensates}
\label{s:discuss}

\subsection{Staggered magnetic domains}

In order to gain a better understanding of the mechanism of the formation
of the staggered magnetic domains shown in Sec.~\ref{s:trap}, we perform
a Bogoliubov analysis of our system.
For simplicity, we assume a homogeneous system with a density of $n_0$,
and set the wave function as $\bm{\psi} = e^{-i \mu t / \hbar}
(\bm{\Psi} + \bm{\phi})$ with $\mu = c_0 n_0$ and $\bm{\Psi} = (0,
\sqrt{n_0}, 0)$, where the three components refer respectively to the
amplitudes of the $m = 1$, $0$, and $-1$ components.
Substituting this into Eqs.~(\ref{GP1}) and (\ref{GP2}) and keeping only
the linear terms with respect to $\bm{\phi}$, we obtain
\begin{subequations}
\begin{eqnarray}
i \hbar \frac{\partial \phi_0}{\partial t} & = & -\frac{\hbar^2}{2 M}
\nabla^2 \phi_0 + c_0 n_0 (\phi_0 + \phi_0^*),
\label{bogo1} \\
i \hbar \frac{\partial \phi_{\pm 1}}{\partial t} & = &
\left( -\frac{\hbar^2}{2 M} \nabla^2 \pm p + q \right) \phi_{\pm 1}
\nonumber \\
& & + c_1 n_0 (\phi_{\pm 1} + \phi_{\mp 1}^*). \label{bogo2}
\end{eqnarray}
\end{subequations}
We can solve these equations by expanding $\bm{\phi}$ as
\begin{equation}
\bm{\phi}(\bm{r}, t) = \sum_{\bm{k}} \left( \bm{u}_{\bm{k}} e^{i
(\bm{k} \cdot \bm{r} - \omega_k t)} + \bm{v}_{\bm{k}}^* e^{-i
(\bm{k} \cdot \bm{r} - \omega_k t)} \right).
\end{equation}
Equation (\ref{bogo1}) has the same form as the equation without the
magnetic field, and the eigenenergy is given by~\cite{Ho,Ohmi}
\begin{equation} \label{omega0}
\hbar \omega_k^0 = \sqrt{\varepsilon_k ( \varepsilon_k + 2 c_0 n_0)},
\end{equation}
where $\varepsilon_k \equiv \hbar^2 k^2 / (2M)$.
The corresponding Bogoliubov eigenvectors $\bm{u}_{\bm{k}}^0$ and
$\bm{v}_{\bm{k}}^0$ are both proportional to $(0, 1, 0)$, which describe
the density modulation in the $m = 0$ component.
It follows from Eq.~(\ref{omega0}) that if $c_0 < 0$, the eigenfrequency
$\omega_k^0$ becomes pure imaginary for long wavelengths, and the system
collapses.

Equation (\ref{bogo2}) has two eigenenergies
\begin{equation} \label{omegapm}
\hbar \omega_k^{\pm} = \pm p + \sqrt{(\varepsilon_k + q) ( \varepsilon_k +
	q + 2 c_1 n_0)},
\end{equation}
and the corresponding Bogoliubov eigenvectors take the forms
\begin{subequations}
\begin{equation} \label{uvp}
\bm{u}_{\bm{k}}^+ \propto (1, 0, 0), \;\;\; \bm{v}_{\bm{k}}^+
\propto (0, 0, 1)
\end{equation}
and
\begin{equation} \label{uvm}
\bm{u}_{\bm{k}}^- \propto (0, 0, 1), \;\;\; \bm{v}_{\bm{k}}^-
\propto (1, 0, 0).
\end{equation}
\end{subequations}
These two modes describe the spin waves that have spin angular momenta
$\pm \hbar$, and the energies are shifted by the linear Zeeman energies
$\pm p$.
The quadratic Zeeman term only shifts the single-particle energy from
$\varepsilon_k$ to $\varepsilon_k + q$.
From Eq.~(\ref{omegapm}), we find that the eigenfrequencies $\omega_k^\pm$
become pure imaginary if $q < 0$ or $2 c_1 n_0 + q < 0$.
Since $c_1 < 0$ and $q > 0$ for the case with the spin-1 ${}^{87}{\rm Rb}$
BEC, the dynamical instabilities in the spin wave occur in a high-density
region.
In such a case, the modes become most unstable when the imaginary part of
$\omega_k^\pm$ becomes maximal, i.e., at wave vectors that meet the
following requirement:
\begin{equation} \label{mostunst}
\varepsilon_k = {\rm max}(0, -c_1 n_0 - q).
\end{equation}

The eigenvectors~(\ref{uvp}) and (\ref{uvm}) have $m = \pm 1$ components,
and the coherent excitation of these modes arises nonzero magnetization in
the $x$-$y$ plane.
Even infinitesimal initial populations in the $m = \pm 1$ components cause
exponential growth if the eigenfrequency is complex, leading to
spontaneous magnetization.
However, if the quadratic Zeeman term is absent (i.e., $q = 0$), the
projected angular momentum on the $x$-$y$ plane $[(F_x^{\rm tot})^2 +
(F_y^{\rm tot})^2]^{1/2}$, as well as $F_z^{\rm tot}$, where $\bm{F}^{\rm
tot} \equiv \int d{\bf r} \bm{F}$, must be conserved, since the linear
Zeeman effect merely rotates the spin and the other terms in the
Hamiltonian are rotation invariant.
In fact, when $q = 0$, the imaginary part of Eq.~(\ref{omegapm}) vanishes
for $k = 0$, which reflects the fact that the uniform magnetization of the
entire condensate is prohibited.
Even in this case, local magnetization is possible, as long as the total
angular momentum is conserved, which leads to the structure formation of
the spin.
Thus, the structure formation of the spin reconciles the spontaneous
magnetization with the spin conservation.
This is a physical account of why the $k \neq 0$ modes in
Eq.~(\ref{omegapm}) become dynamically unstable.

From Eq.~(\ref{mostunst}), we find that the most unstable wavelength in
the situation shown in Fig.~\ref{f:spin} is given by (see also
Refs.~\cite{Mueller,Ueda,Robins})
\begin{equation} \label{lambda}
\lambda = \frac{k}{2\pi} = \frac{2 \pi \hbar}{\sqrt{-2 m (c_1 n_0 + q)}}
\simeq 11 \mu{\rm m}.
\end{equation}
From Fig.~\ref{f:spin} (b), we can see that the size of the staggered
domains around $z \sim 0$ is roughly equal to $13 \mu{\rm m}$, which is in
reasonable agreement with Eq.~(\ref{lambda}).
The size increases with $|z|$, since the density decreases with $|z|$.
The wavelength (\ref{lambda}) is also consistent with the interval between
the concentric rings $\simeq 20 \mu{\rm m}$ shown in Fig.~\ref{f:2d}.

We note that the spin domains observed in Ref.~\cite{Miesner} may also be
regarded as a consequence of structure formation due to spin
conservation.
In Ref.~\cite{Miesner}, the spin-1 ${}^{23}{\rm Na}$ BEC is prepared at
a 1:1 mixture of the $m = 0$ and $m = 1$ states; then the length of the
initial spin vector is $\sqrt{3} / 2$.
Due to the antiferromagnetic interaction, the spin vector tends to
vanish, whereas the $z$ component of the total spin must be conserved.
(The $x$ and $y$ components need not be conserved because of the presence
of magnetic field in the $z$ direction.)
In the cigar-shaped trap of Ref.~\cite{Miesner}, the system thus responds
to form staggered magnetic domains.
In fact, the $z$ component of the total spin is conserved, whereas the
average length of the spin vector decreases from $\sqrt{3} / 2$ to $1 / 2$
by the domain formation.

\subsection{Helical structure}

The spin vector of the helical structure in Fig.~\ref{f:vector} (d) can be
written as $(e^{-i \alpha z}, \sqrt{2}, e^{i \alpha z}) / 2$, where $2\pi
/ \alpha$ is a pitch of the helix.
The torsion energy per atom is then given by $\hbar^2 \alpha^2 / (4 m)$,
and the ferromagnetic-interaction energy is given by $c_1 n / 2$.
In order to compare the energy of the helical structure with that of the
staggered domain structure, we assume that the spin vector of the latter
varies in space as $(\sin \frac{\alpha z}{2}, \sqrt{2} \cos \frac{\alpha
z}{2}, \sin \frac{\alpha z}{2}) / \sqrt{2}$.
The kinetic and ferromagnetic-interaction energies of this state become
$\hbar^2 \alpha^2 / (8 m)$ and $c_1 n / 4$, respectively.
Therefore, the helical structure is energetically more favorable than the
domain structure when $|c_1 n| > \hbar^2 \alpha^2 / (2 m)$.
In the present case, $\hbar^2 \alpha^2 / (2 m k_{\rm B}) \sim 0.2$ nK
and $c_1 n / k_{\rm B} \sim 1$ nK, and hence the change from the domain
structure to the helical structure is favored.
Since the energy is conserved in the present situation, the excess energy
associated with this structure change is considered to be used to excite
the helical state.
Thus the helix in Fig.~\ref{f:vector} (d) should be regarded as being in
an excited state.
This explains why the helical structure appears only transiently in the
present situation.
If the excess energy can be dissipated by some means, the helical
structure is expected to have a longer lifetime.

In solid state physics, there are two types of magnetic domain walls: the
Bloch wall and the N\'eel wall.
At the Bloch wall spin flip occurs by tracing a helix, while at the
N\'eel wall the spin flip occurs in a plane~\cite{Kittel}.
However, the domain walls in Fig.~\ref{f:vector} (c) are categorized as
neither type of wall, since the spin vector vanishes in the middle of the
wall.
In the situation given in Fig.~\ref{f:vector}, the formation of the
staggered magnetic domains in the $x$ direction is followed by the growth
of the $y$ component, leading to the formation of the helical structure.
The helical structure may therefore be regarded as the formation of
Bloch walls.
On the other hand, N\'eel walls may also be formed, depending on the
initial conditions.

\subsection{Magnetic field dependence}

The above calculations were carried out for the magnetic field of $B = 54$
mG, which gives $q / |c_1 n| \sim 0.01$, so the effect of the magnetic
field was very small.
To investigate the magnetic field dependence of the structure formation of
the spin, we performed numerical simulations for various strengths of the
magnetic field.
The spin domain structure for $B = 0$ is qualitatively similar to that
for $B = 54$ mG.
However, the initial, large-scale spin exchange shown in
Fig.~\ref{f:evolution} (b) ($\omega_\perp t = 300$) is absent in the case
of $B = 0$, since the imaginary part of $\hbar \omega_k^{\pm}$ in
Eq.~(\ref{omegapm}) for long wavelengths, $\simeq (q |q + 2 c_1
n|)^{1/2}$, vanishes when $q = 0$.
The imaginary part for long wavelengths increases with the strength of
the magnetic field for $B \lesssim 500$ mG.
In fact, when $B = 200$ mG, which gives $q / |c_1 n| \sim 0.1$, the
initial spin exchange is enhanced compared with the case when $B = 54$
mG.
For $B = 200$ mG, the size of the spin domain becomes larger, in agreement
with the fact that Eq.~(\ref{lambda}) gives $\lambda \simeq 16 \mu {\rm
m}$.
When $B = 1$ G ($q / |c_1 n| \simeq 4$), the eigenfrequencies
$\omega_k^{\pm}$ are always real, and no spin dynamics occur.
This is because for large $B$, the $m = 0$ state becomes the ground state
in the subspace of $F^{\rm tot}_z = 0$, due to the quadratic Zeeman
effect.

\subsection{Initial conditions}

\begin{figure}
\includegraphics[width=8.4cm]{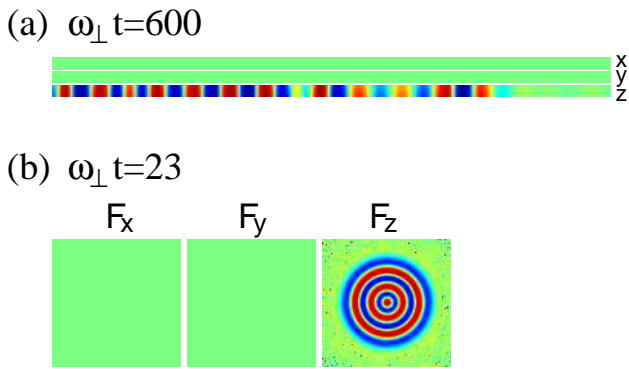}
\caption{
The three components of the mean spin vector $F_x / n$, $F_y / n$, $F_z /
n$ (from top to bottom panel in (a) and from left to right panel in (b))
for the initial state $\psi_1 = \sqrt{1 + 0.01} \psi_{\rm g}$, $\psi_0 =
0$, and $\psi_{-1} = \sqrt{1 - 0.01} \psi_{\rm g}$, where $\psi_{\rm g}$
is the $m = -1$ ground state.
The other conditions are the same as those in Fig.~\protect\ref{f:spin}
for (a) and in Fig.~\ref{f:2d} for (b).
}
\label{f:spin2}
\end{figure}

We have investigated the structure formation of the spin for several other
initial states.
For example, Fig.~\ref{f:spin2} shows the structure formation obtained for
the initial state $\psi_1 = \sqrt{1 + 0.01} \psi_{\rm g}$, $\psi_0 = 0$,
and $\psi_{-1} = \sqrt{1 - 0.01} \psi_{\rm g}$, where $\psi_{\rm g}$ is
the $m = -1$ ground state.
Spontaneous magnetization from this initial state also proceeds with the
formation of the staggered domain structure for the cigar-shaped geometry,
and the concentric ring structure for the pancake-shaped geometry.
The magnetization occurs in the $z$ direction, since the initial spin has
a small $z$ component.
Thus, we can conclude that the spatial spin structure formed in the
spontaneous magnetization is essentially determined by the geometry of the
system and does not depend on the initial state.

\section{Conclusions}
\label{s:conclusions}

We have studied the spontaneous local magnetization of the ferromagnetic
spin-1 condensate that is initially prepared in the $m = 0$ state and
subject to spin conservation.
An initial infinitesimal magnetization of one component gives rise to the
exponential growth of that component due to the dynamical instabilities.
As a consequence, the staggered spin domain structure is formed along the
trap axis in the case of an elongated cigar-shaped trap, where the local
mean spin vector undergoes the Larmor precession in the presence of a
magnetic field.
The size of the domain structure agrees with that obtained by Bogoliubov
analysis (Eq.~(\ref{lambda})).
The helical structures also appear transiently.
In the case of a tight pancake-shaped trap, a concentric ring structure
is formed.

The underlying physics of the structure formation in the magnetization is
the competition between the ferromagnetic spin correlation and the
conservation of the total magnetization.
The magnetization of the entire system in the same direction is
prohibited.
As a consequence of the competition, magnetization occurs only locally,
resulting in spatial structures.
In solid state physics, in contrast, the magnetization of an entire system
is possible because neither the energy nor the angular momentum is
conserved, due to the interaction of the system with its environment.

In this paper, we have considered only elongated cigar-shaped geometry
and tight pancake-shaped geometry.
In other systems, such as the isotropic system and the rotating system
with vortices, other intriguing spin structures or spin textures are
expected to emerge in a spontaneous manner, because these systems involve
dynamical instabilities of different symmetries.
We plan to report the results of these cases in a future publication.

\begin{acknowledgments}
We would like to thank S. Inouye for helpful discussions.
This work was supported by Special Coordination Funds for Promoting
Science and Technology, by a 21st Century COE program at Tokyo Tech
``Nanometer-Scale Quantum Physics,'' and by a Grant-in-Aid for Scientific
Research (Grant No. 15340129) from the Ministry of Education, Science,
Sports, and Culture of Japan.
\end{acknowledgments}

\end{document}